\documentclass[8.5pt,twoside,twocolumn]{article}
\usepackage[super,sort&compress,comma]{natbib}
\usepackage{mhchem}
\usepackage{times,mathptmx}
\usepackage{sectsty}
\usepackage{balance}

\usepackage{graphicx} 
\usepackage{epsfig}
\usepackage{lastpage}
\usepackage[format=plain,justification=raggedright,singlelinecheck=false,font=small,labelfont=bf,labelsep=space]{caption}
\usepackage{fancyhdr}
\begin{document}

\thispagestyle{plain}
\fancypagestyle{plain}{
5\renewcommand{\headrulewidth}{1pt}}
\setcounter{secnumdepth}{5}

\makeatletter
\def\subsubsection{\@startsection{subsubsection}{3}{10pt}{-1.25ex plus -1ex minus -.1ex}{0ex plus 0ex}{\normalsize\bf}}
\def\paragraph{\@startsection{paragraph}{4}{10pt}{-1.25ex plus -1ex minus -.1ex}{0ex plus 0ex}{\normalsize\textit}}
\renewcommand\@biblabel[1]{#1}
\renewcommand\@makefntext[1]%
{\noindent\makebox[0pt][r]{\@thefnmark\,}#1}
\makeatother
\renewcommand{\figurename}{\small{Fig.}~}
\sectionfont{\large}
\subsectionfont{\normalsize}


\twocolumn[
 \begin{@twocolumnfalse}
\noindent\LARGE{\textbf{Advanced Materials for Solid-State Refrigeration.}}
\vspace{0.6cm}

\noindent\large{\textbf{Llu\'{\i}s Ma\~nosa,$^{\ast}$\textit{$^{a}$} Antoni Planes,\textit{$^{a}$} and
Mehmet Acet\textit{$^{b}$}}}\vspace{0.5cm}

\noindent\textit{\small{\textbf{Received Xth XXXXXXXXXX 20XX, Accepted Xth XXXXXXXXX 20XX\newline
First published on the web Xth XXXXXXXXXX 200X}}}

\noindent \textbf{\small{DOI: 10.1039/b000000x}}
\vspace{0.6cm}

\noindent \normalsize{Recent progress on caloric effects are reviewed. The application of external stimuli such as magnetic field, hydrostatic pressure, uniaxial stress and electric field give rise respectively to magnetocaloric, barocaloric, elastocaloric and electrocaloric effects. The values of the relevant quantities such as isothermal entropy and adiabatic temperature-changes are compiled for selected materials. Large values for these quantities are found when the material is in the vicinity of a phase transition. Quite often there is coupling between different degrees of freedom, and the material can exhibit cross-response to different external fields. In this case, the material can exhibit either conventional or inverse caloric effects when a field is applied. The values reported for the many caloric effects at moderate fields are large enough to envisage future application of these materials in efficient and environmental friendly refrigeration.}
\vspace{0.5cm}
 \end{@twocolumnfalse}
 ]

\section{Introduction}


\footnotetext{\textit{$^{a}$~Departament Estructura i Constituents de la Mat\`eria, Facultat de F\'{\i}sica.
Universitat de Barcelona. Diagonal 645. 08028 Barcelona. Catalonia. Fax: 34 934037063; Tel: 34 934039181; E-mail: lluis@ecm.ub.edu; toni@ecm.ub.edu}}
\footnotetext{\textit{$^{b}$~Experimentalphysik. Universit\"at Duisburg-Essen D-47048 Duisburg. Germany. E-mail: mehmet.acet@uni-due.de.}}



Every material changes its temperature when subjected to a sudden change of an external field (electric, mechanical, magnetic...). This property is generally known as the caloric effect and is related to a change in the material's entropy when the external field is isothermally modified. Generally, around room temperature, the magnitude of such a caloric effect is small and the temperature-change only becomes relatively large at very low-temperatures when the specific heat of the material is low. For this reason, refrigeration based on sweeping an external field adiabatically has been limited to low-temperatures for very long. The typical example is the attainment of cryogenic temperatures by making use of the magnetocaloric effect in paramagnetic salts when adiabatically demagnetized \cite{Giauque1935}.

The efficient use of the magnetocaloric effect for refrigeration around ambient temperature was proposed by Brown in the late seventies, \cite{Brown1976} and a major breakthrough took place at the end of the nineties with the discovery by Pecharsky and Gschneidner of a material exhibiting large magnetic-field-induced entropy and temperature-changes around room temperature; the intermetallic Gd-Si-Ge compound, which was denoted as a giant magnetocaloric material \cite{Pecharsky1997}. This discovery boosted the research in the field, and nowadays there is a large variety of materials which have been reported to exhibit the giant magnetocaloric effect. The key feature for the effect to be giant is the presence of a first-order phase transition that most of these materials undergo. It is the transition entropy-change (associated with the latent heat of a first-order phase transition) that makes the major contribution to the field-induced entropy-change of the magnetocaloric effect. As a consequence of strong interplay between magnetic and structural degrees of freedom, the phase transition also involves typically a change in the crystallographic structure.

\begin{figure}[h]
\centerline{\includegraphics[height=8cm]{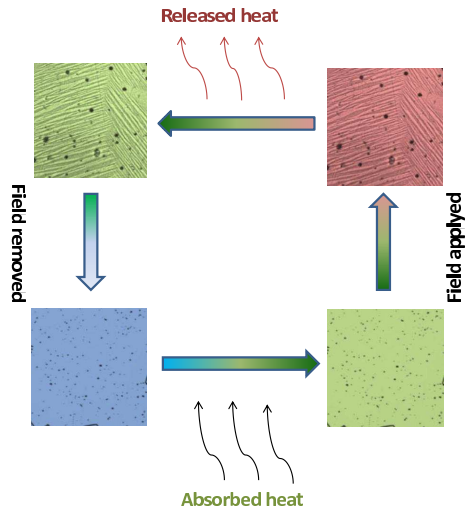}}
 \caption{Refrigerating cycle. Schematic diagram for a solid-state based refrigerating cycle based on a conventional giant caloric effect. On the first stage the field is adiabatically applyied resulting in a temperature temperature increase of the material (a forward phase transition occurs in this stage). On the second stage the material is allowed to cool down at constant field by transferring heat to a heat sink. On the third stage the field is adiabatically removed and the material cools further (the reverse phase transition occurs in this stage). In the fourth stage, the cool material absorbs heat from the cold reservoir (which becomes colder) and recovers the initial state. The images to illustrate the cycle correspond to a magnetic shape memory alloy which undergoes a martensitic phase transition. The field can be either magnetic field, hysdrostatic pressure or uniaxial stress.}
 \label{Refrigeratingcycle}
\end{figure}

While most of the efforts are nowadays devoted to investigating the magnetocaloric effect, the possibility of inducing the phase transition in the solid state by fields other than magnetic opens up new routes for solid-state refrigeration using many different external stimuli. Indeed, materials showing significant changes in a certain thermodynamic quantity at the phase transition $-$ such as volume, strain or polarization $-$ will be extremely sensitive to the application of the corresponding thermodynamically conjugated field $-$ pressure, stress and electric field, respectively. Hence, the possibility of inducing the transition by means of pressure, uniaxial stress and electric field will give rise to the barocaloric, elastocaloric and electrocaloric effects, respectively. A solid-state cooling cycle is illustrated in Fig. \ref{Refrigeratingcycle} based on a giant caloric effect attained by applying a generic external field.

Usually the material's entropy decreases when an external field is applied isothermally, and its temperature increases when a field is applied adiabatically. Since this is the most frequent situation, the caloric effect is known as 'conventional'. However, there are some cases for which the situation is reversed: the entropy increases on applying a field isothermally, and the material cools. This effect is known as the 'inverse' caloric effect. Inverse caloric effects are possible in those systems with a strong coupling between different degrees of freedom, with a cross-response to the external fields, and for which the entropy contains contributions from all these degrees of freedom.

There have been detailed and comprehensive reviews dealing with magnetocaloric materials \cite{Gschneidner2005,Bruck2005,Bruck2008,Planes2009,Shen2009,deOliveira2010,Smith2012,Franco2012} and other caloric effects \cite{Lu2009,Fahler2012,Scott2011,Valant2012,Planes2012}. Also, details on the chemistry of the different materials illustrated in the present paper can be found in 
\cite{Gschneidner2005,Franco2012,Lines1977,Otsuka1998,Graf2011}, and references therein. It is not intended to duplicate that effort here. The present paper is aimed at presenting the different caloric effects reported so far. For each of them, we select a particular material which conveniently illustrates the case.



\section{Thermodynamics of caloric effects.}

Let a generic system in thermodynamic equilibrium be described in terms of the generalized displacements, $x_i$, forces $Y_i$ (fields) and temperature $T$. Both displacements and forces have the same tensorial character. However, a scalar approach is appropriate to describe most cases of interest (where only the magnitude of the external field is varied), and this will be considered in the following. The change in entropy reads:

\begin{equation}
dS= \frac{C}{T} dT + \sum_i \left( \frac {\partial x_i}{\partial T} \right)_{Y_{j}} dY_i ,
\label{generaldS}
\end{equation}

\noindent where we have made use of the Maxwell relations $\left( \frac {\partial S}{\partial Y_i} \right)_{T,Y_{j \neq i}} = \left( \frac {\partial x_i}{\partial T} \right)_{Y_{j}}$, and $C$ is the heat capacity.

For an isothermal change of a given field from 0 to $Y$, the field-induced entropy-change accounting for the caloric effect is given by:

\begin{equation}
S(T,Y)-S(T,Y=0)=\Delta S(T,Y)= \int_0^{Y} \left( \frac {\partial x}{\partial T} \right)_{Y} dY.
\label{caloricDeltaS}
\end{equation}

When the field is applied adiabatically, the corresponding change in temperature is given by:

\begin{equation}
\Delta T = - \int_0^{Y} \frac{T}{C} \left( \frac {\partial x}{\partial T} \right)_{Y} dY
\label{caloricDeltaT}
\end{equation}

The above expressions quantify the magnetocaloric ($Y=H$ and $x=M$), barocaloric ($Y=-p$ and $x=V$), elastocaloric ($Y=\sigma$ and $x=\epsilon$) and electrocaloric ($Y=E$ and $x=P$) effects, where $H$ is the magnetic field, $M$, magnetization, $p$, hydrostatic pressure, $V$ volume, $\sigma$, uniaxial stress, $\epsilon$, uniaxial strain, $E$, electric field and $P$, polarization.

\subsection{Experimental determination of the caloric effects}

The most commonly used method for the determination of the entropy-change makes use of isothermal $x$ vs $Y$ curves to numerically integrate eq. \ref{caloricDeltaS}:

\begin{equation}
\Delta S \left( Y,T(k) \right)= \frac{1}{\Delta T_k} \left[ \int_0^Y x(T_{k+1}) dY - \int_0^Y x(T_{k}) dY \right]
\label{numericalintegrationMaxwell}
\end{equation}

\noindent where $T(k) = (T_{k+1}+T_k)/2$, $\Delta T_k= T_{k+1}-T_k$. While this provides a fast and easy procedure to quantify a caloric effect, there has been considerable controversy on whether or not the method can be applied to first-order phase transitions \cite{Giguere1999,Sun2000,Casanova2002a,Zhou2009,Manosa2009}. Large spurious values \cite{deCampos2006} were shown to arise from inappropriate experimental protocols \cite{Liu2007}. The major issue associated with this problem is the existence of a hysteresis at the first-order phase transition. However, at present, it is clear that the method gives reliable data for $\Delta S$, provided that the experimental protocol is the appropriate one which correctly takes into account the hysteresis of the transition (details can be found in refs. \cite{Caron2009,Planes2009,Tocado2009}). Alternatively, $\Delta S$ can also be obtained by numerically integrating the derivatives $\left( \frac {\partial x}{\partial T} \right)_{Y}$ computed from measured isofield $x(T)$ curves. The latter method ensures that the sample completes the full phase transformation and consequently is free from giving spurious values.

An alternative way to obtain $\Delta S$ is from calorimetric measurements of the temperature dependence of the specific heat of the material for different constant values of the external field $Y$ (see equation \ref{generaldS}). As previously mentioned, most giant caloric materials undergo a first-order phase transition, and the best suited calorimetric technique for first-order phase transitions is differential scanning calorimetry (DSC). Purpose-built DSC operating under different external fields have been developed to study several caloric effects \cite{Plackowski2002,Marcos2003,Guyomar2006,Basso2008,Moya2012}. These devices provide heat flux as a function of temperature, from which the entropy (referenced to a given state at $T_0$) is obtained as:

\begin{equation}
S(T,Y)-S(T_0,Y) = \int_{T_0}^T \frac{1}{T} \frac{\dot{Q}(Y)}{\dot{T}} dT
\label{integratedcalorimetryT}
\end{equation}

\noindent where $\dot{Q}$ and $\dot{T}$ are the heat flux and cooling (or heating) rate, respectively. The field-induced entropy-change (accounting for the caloric effect) is readily obtained by subtracting the integrated curves resulting from equation \ref{integratedcalorimetryT}.

Some of these devices can also operate at constant temperature and sweeping the external field. This method provides a direct determination of the field-induced entropy-change which is given by:

\begin{equation}
S(T,Y)-S(T,0) = \frac{1}{T} \int_{0}^Y \frac{\dot{Q}(T)}{\dot{Y}} dY
\label{integratedcalorimetryY}
\end{equation}

\noindent where $\dot{Y}$ is the field rate.

For a first-order phase transition, the shift in the transition temperature with field is governed by the Clausius-Clapeyron equation:

\begin{equation}
\frac{dT}{dY} = - \frac{\Delta x}{\Delta S}
\label{clausiusclapeyron}
\end{equation}

\noindent Since $\Delta S$ is negative (lower entropy of the low-temperature phase), the shift in the transition is determined by the sign of the generalized displacement discontinuity, $\Delta x$.

Finally, a direct determination of a caloric effect is also achieved by measuring the temperature of the sample (using an appropriate thermometer) while the external field is adiabatically modified. Although this is a quite direct method to quantify a caloric effect, adiabatic temperature-change experiments are usually difficult to implement, and the majority of studies are devoted to field-induced entropy-changes.

\section{Gd-Si-Ge compounds. The giant prototype magnetocaloric material.}

At room temperature Gd$_5$Si$_2$Ge$_2$ is paramagnetic and exhibits a monoclinic structure (space group $P112_1/a$). On cooling it undergoes a magnetostructural transition to an orthorhombic (space group $Pnma$) structure which is ferromagnetically ordered. Since Gd$_5$Ge$_4$ is antiferromagnetic and Gd$_5$Si$_4$ is ferromagnetic, the phase diagram for off-stoichiometric Gd$_5$(Ge$_{1-x}$Si$_x$)$_4$ is very rich. For all $x$, the ground state is ferromagnetic with an orthorhombic structure (space group $Pnma$), but different structural and magnetic phases can be found depending on $x$. A detailed phase diagram is described in \cite{Gschneidner2005,Bruck2005}. The giant magnetocaloric effect is observed for alloys with $0.3\leq x\leq 0.5$. It is within this composition region that the alloy undergoes the paramagnetic to ferromagnetic first-order magnetostructural transition with a transition temperature that linearly increases with increasing Si concentration.

The structural transition involves shear displacements of atomic layers in which (Si,Ge)-(Si,Ge) dimers that are richer in Ge increase their distances. This modifies the spin-dependent hybridization between Ge $4p$ and Gd $5d$ conduction states reducing the net Gd 5$d$ moment and the strength of the ferromagnetic RKKY exchange coupling across sheared layers \cite{Haskel2007,Paudyal2006}. In addition to the giant magnetocaloric effect, the magnetostructural transition is also at the origin of other technologically important properties of this alloy, such as giant magnetoresistance \cite{Morellon1998b} and magnetostriction \cite{Morellon1998a}.

\begin{figure}[h]
\centerline{\includegraphics[height=7cm]{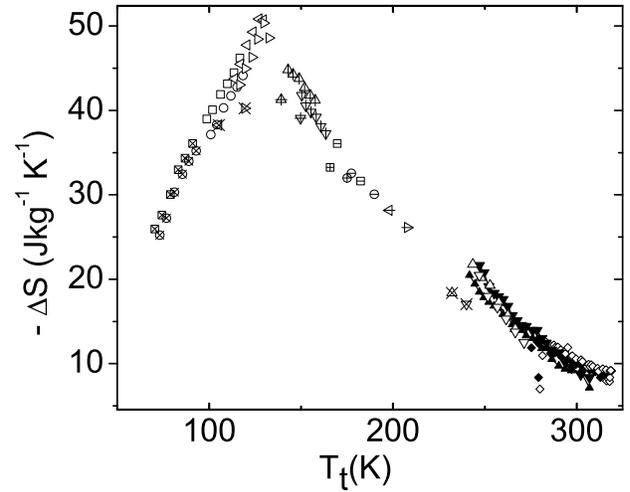}}
 \caption{Magnetocaloric effect in Gd-Si-Ge. Transition entropy-change as a function of the transition temperature for
 Gd$_5$(Ge$_{1-x}$Si$_x$)$_4$ alloys. Figure adapted from ref. \cite{Marcos2002}}
 \label{figDeltaSGdSiGe}
\end{figure}

The magnetocaloric effect in the Gd$_5$(Ge$_{1-x}$Si$_x$)$_4$ family has been investigated by numerous researchers by means of several experimental techniques. Depending on composition and purity of the samples, the reported entropy values range from 10 Jkg$^{-1}$K$^{-1}$ up to 50 Jkg$^{-1}$K$^{-1}$. The highest values correspond to those compositions for which the magnetostructural transition is close to the Neel temperature of the orthorhombic phase. Interestingly, the whole entropy-change has been found to scale with the transition temperature, no matter if the latter is tuned by composition or magnetic field \cite{Casanova2002a}. This result is illustrated in fig. \ref{figDeltaSGdSiGe} and proves that magnetovolume effects due to the magnetic field are of the same nature as the volume effects caused by substitution. Such a statement has been corroborated by later studies showing the equivalence of temperature, magnetic field, and chemical and hydrostatic pressures on the polymorphism of Gd$_5$(Ge$_{1-x}$Si$_x$)$_4$ \cite{Casanova2004,Mudryk2005}.

\begin{figure}[h]
\centerline{\includegraphics[height=7cm]{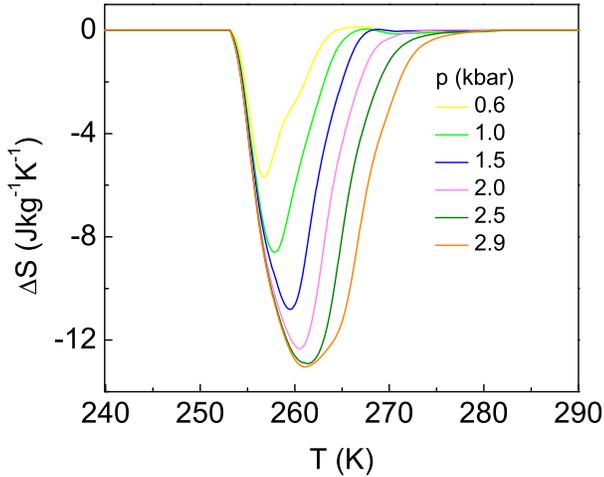}}
 \caption{Barocaloric effect in Gd-Si-Ge. Temperature dependence of the pressure induced entropy-change for
 Gd$_5$Si$_2$Ge$_2$. Figure adapted from ref. \cite{Yuce2012}.}
 \label{figbarocaloricGdSiGe}
\end{figure}

The magnetostructural transition involves a change in volume of the unit cell: the volume of the low-temperature monoclinic phase is larger than that of the ferromagnetic orthorhombic one. Owing to such a volume change, the transition is sensitive to external hydrostatic pressure, and therefore, it is prone to exhibit a giant barocaloric effect. Such a possibility was theoretically predicted within the framework of band theory \cite{Medeiros2008}, and experimentally demonstrated by differential scanning calorimetry under hydrostatic pressure \cite{Yuce2012}. Figure \ref{figbarocaloricGdSiGe} shows the pressure-induced entropy-change as a function of temperature for selected values of hydrostatic pressure for stoichiometric Gd$_5$Si$_2$Ge$_2$. The values found for moderate pressures around 3 kbar are comparable to those corresponding to the magnetocaloric effect for fields around 2 T for this compound.

For Gd$_5$Si$_2$Ge$_2$, both magnetocaloric and barocaloric effects are found to be conventional, i.e. the entropy decreases with increasing field and pressure. This is associated to the larger magnetization and lower volume of the low-temperature ferromagnetic orthorhombic phase. The total entropy of the alloy decreases at the paramagnetic-monoclinic to ferromagnetic-orthorhombic transition, and such a decrease is due to a decrease in both magnetic and structural contributions to the entropy.

\section{La-Fe-Si family and the inverse barocaloric effect.}

The La(Fe$_{13-x}$Si$_x$) compounds have a cubic NaZn$_{13}$ structure (space group $Fm\bar{3}c$) in the concentration range 1.2$\leq x \leq$2.5. La atoms occupy the 8a sites. The 8b site is fully occupied by Fe, and the 96i site is shared by Fe and Si atoms \cite{Bruck2005}. On cooling, the alloy orders ferromagnetically at a Curie temperature ($T_C$) that increases from 198 K for $x=1.5$ to 262 K for $x=2.5$. The saturation magnetic moment decreases from 2.08 $\mu_B$ to 1.85 $\mu_B$ in this range \cite{Palstra1983}. Above $T_C$, an itinerant-electron metagenetic (IEM) transition can be induced by an external magnetic field \cite{Fujita2003}. The transition has a marked first-order character for low $x$, but for $x\geq$2.4, it becomes second order.

In contrast to Gd$_5$Si$_2$Ge$_2$, the phase transition does not involve structural symmetry breaking, and the material keeps the cubic $Fm\bar{3}c$ structure above and below the transition temperature. However, a strong coupling between magnetism and structure is evidenced by the change in the unit cell volume: the volume of the low-temperature ferromagnetic phase is $\sim$ 1 \% larger than that of the high-temperature paramagnetic phase \cite{Fujita2001}. These large magnetization and volume changes at the phase transition suggests that this alloy might have interesting magnetocaloric and barocaloric properties.

The IEM trasition originates from a magnetic-field-induced change in the density of states of the 3d electrons at the Fermi level \cite{Fujita2003}. A strong magnetovolume effect originates from the existence of a peak in the electronic density of states close to the Fermi level that yields a negative contribution from spin fluctuations to the magnetostriction, resulting in a volume increase on cooling from the paramagnetic to the ferromagnetic state.

From the point of view of potential applications, La(Fe$_{13-x}$Si$_x$) is an attractive material due to its reduced hysteresis at the first-order phase transition (irreversibilities related to hysteresis reduce the cooling capacity of the alloy). It only has the drawback that the transition temperature is below room temperature. Two alternative methods have been reported to bring the transition temperature to values close to room temperature \cite{Shen2009}. One method is to add interstitial elements (H is the most successful one) that expand the crystalline lattice and modify the exchange interaction between iron atoms. The electron band-structure is not modified but the tri-critical point is shifted to higher temperatures as a consequence of the change in the lattice parameter. Another method is to replace some Fe atoms by other magnetic transition metals (Co has been found to be the best suited element). Such a substitution does modify the electron band-structure and consequently the relative phase stability. As the amount of Co increases, the first-order character of the transition weakens; and for high enough Co content, the phase transition turns to second order. \cite{Liu2004}

\begin{figure}[h]
\centerline{\includegraphics[height=6cm]{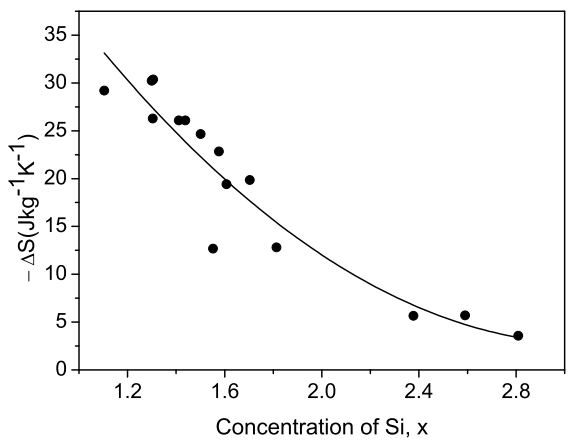}}
 \caption{Magnetocaloric effect in La-Fe-Si. Magnetic-field entropy-change as a function of the Si concentration for La(Fe$_{13-x}$Si$_x$). Figure adapted from ref. \cite{Gschneidner2005}. An average density of 7229 kgm$^3$ has been used for all data. The line is a guide to the eyes.}
 \label{figDeltaSLaFeSi}
\end{figure}

The study of the magnetocaloric effect in La(Fe$_{13-x}$Si$_x$) has received considerable interest \cite{Shen2009}. Figure \ref{figDeltaSLaFeSi} shows reported values for the isothermal entropy-change as a function of Si concentration. A marked decrease is observed as $x$ increases: The first-order character of the transition diminishes and approaches a second order character. Such a decrease is overcome by tailoring the transition temperature by the concentration of interstitial hydrogen in hydrogenated La(Fe$_{13-x}$Si$_x$)H$_y$ compounds. While $T_c$ linearly rises with increasing $y$, the entropy-change only exhibits a very weak dependence on the hydrogen content \cite{Chen2003,Fujita2003}.

\begin{figure}[h]
\centerline{\includegraphics[height=7cm]{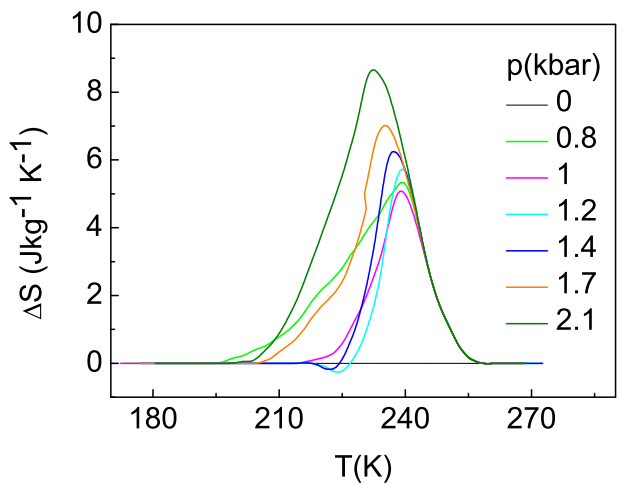}}
 \caption{Barocaloric effect in La-Fe-Si. Temperature dependence of the pressure-induced entropy-change for
 LaFe$_{11.33}$Co$_{0.47}$Si$_{1.2}$. Figure adapted from ref. \cite{Manosa2011}.}
 \label{figbarocaloricLaFeSi}
\end{figure}

The pressure dependence of the magnetic and magnetocaloric properties of La-Fe-Si and doped compounds has been reported by several experimental techniques, and it has also been theoretically modeled \cite{Fujita2003,Lyubina2008}. The barocaloric effect in LaFe$_{11.33}$Co$_{0.47}$Si$_{1.2}$ was measured by means of differential scanning calorimetry under applied hydrostatic pressure \cite{Manosa2011}. Figure \ref{figbarocaloricLaFeSi} shows the values obtained for the pressure-induced entropy-change as a function of temperature for selected values of the hydrostatic pressure. The barocaloric effect increases with pressure, and for pressures $\sim$2 kbar, the barocaloric effect amounts to about 75\% of the transition entropy-change. This value compares well with the magnetocaloric effect for a magnetic field of $\sim$5T for the same compound. The most striking feature in fig. \ref{figbarocaloricLaFeSi} is the positive value for the entropy-change which indicates that the barocaloric effect in this material is inverse: the entropy increases with increasing pressure. This finding is consistent with the decrease in the temperature of the magnetostructural transition with increasing pressure.

\begin{figure}[h]
\centerline{\includegraphics[height=6cm]{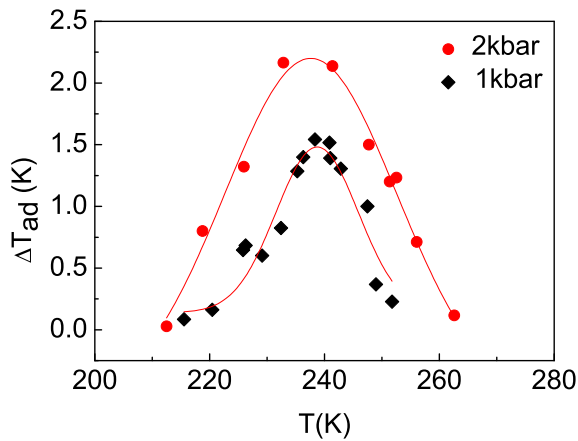}}
 \caption{Barocaloric effect in La-Fe-Si. Temperature dependence of the adiabatic temperature-change for decompression of 2 kbar (circles) an 1kbar (diamonds) for LaFe$_{11.33}$Co$_{0.47}$Si$_{1.2}$. Figure adapted from ref. \cite{Manosa2011}.}
 \label{figDeltaTLaFeSi}
\end{figure}

A material exhibiting the inverse barocaloric effect will have the unusual property of cooling when adiabatically compressed and warming when decompressed. Such a feature is illustrated in figure \ref{figDeltaTLaFeSi} which shows measurements of the adiabatic temperature-change when a LaFe$_{11.33}$Co$_{0.47}$Si$_{1.2}$ is rapidly decompressed. The positive values are direct evidence of the inverse nature for the barocaloric effect in this compound.

In systems with coupling between different degrees of freedom, the dominant change in entropy at the transition should be associated with a conventional caloric effect, whereas the secondary property may provide conventional or inverse effect depending of the specific feature of the coupling. In La-Fe-Si the magnetocaloric effect is conventional while the barocaloric is inverse, which reflects that the transition is essentially of magnetic nature. Indeed the major contribution to the entropy in these compounds is due to the magnetic contribution from itinerant $3d$ electrons. Spin fluctuations induce a larger volume of the ferromagnetic phase, and as a consequence, the barocaloric effect in this material is inverse.

\section{Thermoelastic shape memory alloys and the elastocaloric effect.}

Shape memory materials have received considerable attention both from the fundamental point of view and from the viewpoint of technological applications of the shape memory effect. They are cubic intermetallics with an open structure (typically $Fm3m$ or $Pm3m$) at high-temperature which upon cooling transform martensitically to a more compact structure with lower symmetry. They are capable of recovering from very large deformations ($\sim$ 10\%) by simply changing its temperature (shape memory effect), or after suitable themo-mechanical training, they shift from one shape to another by warming and cooling (two-way shape memory effect). They also exhibit superelastic behaviour, related to the recovery of large strains upon loading and unloading the sample. These peculiar mechanical properties are related to the martensitic transition that these alloys undergo. The transition is first-order from a high-temperature high-symmetry phase to a low-temperature lower symmetry phase where the lattice distortion can be mainly described by a shear mechanism. An excellent overview of these materials can be found in \cite{Otsuka1998}.

In non-magnetic shape memory alloys, there is negligible difference in the volume of the unit cell of the martensitic and cubic phases. Nevertheless, the transformation involves a large shear distortion (typically shearing \{110\} planes along the $\langle 1 \bar{1} 0 \rangle$ directions), and therefore, it is very sensitive to the application of uniaxial stresses. Indeed the possibility of inducing martensitic transitions by mechanical stress has been known since decades, and many of the mechanical applications of these alloys rely on such a possibility.

\begin{figure}[h]
\centerline{\includegraphics[height=6cm]{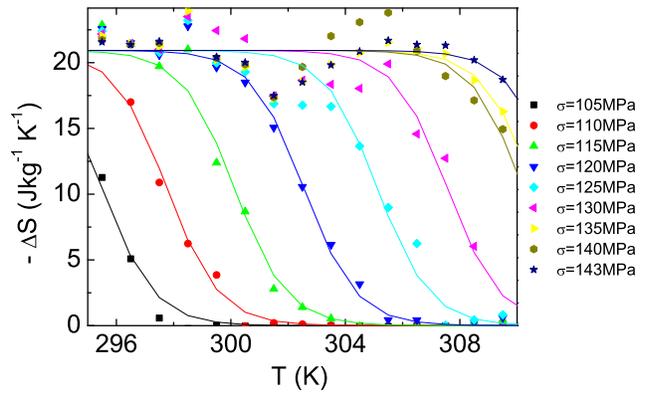}}
 \caption{Elastocaloric effect in Cu-Zn-Al. Temperature dependence of the stress-induced entropy-change for Cu$_{68.13}$Zn$_{15.74}$Al$_{16.13}$. Figure adapted from ref. \cite{Bonnot2008}.}
 \label{figDeltaSelastocaloric}
\end{figure}

The entropy-change associated with the latent heat of the first-order martensitic transition gives rise to a large caloric effect when the transition is induced by uniaxial stress: This is the elastocaloric effect. A giant elastocaloric effect was reported for a Cu-Zn-Al single crystal \cite{Bonnot2008} and for Ti-Ni polycrystals \cite{Cui2012,Bechtold2012}. It is also worth mentioning that the elastocaloric effect was earlier reported for non martensitic Fe-Rh alloys \cite{Annaorazov1996}. The temperature-dependence of the stress-induced entropy-change (elastocaloric effect) for Cu-Zn-Al is illustrated in figure \ref{figDeltaSelastocaloric} for selected values of applied stress. There are a few salient features in comparison to the other caloric effects referred in the previous sections. First of all, since the sample is ductile and the transition temperature is very sensitive to stress, it is possible to induce the full transformation in the whole sample, and consequently, the stress-induced entropy values coincide with the total transition entropy-change. A second feature is that the transition can be shifted to temperatures well above the zero-stress transition temperatures. This results in a large plateau in $\Delta S$ (fig. \ref{figDeltaSelastocaloric}) which reflects that such a giant caloric effect spans over a broad temperature range giving rise to a very large refrigerant capacity in these materials. We note here that the data shown in fig. \ref{figDeltaSelastocaloric} correspond to an alloy with a transition temperature (in absence of stress) well below room temperature (T$_M$=234K), and experiments carried out close to this temperature show that stresses as low as 5 MPa are enough to induce the transformation in the whole sample.

The elastocaloric effect in shape memory alloys is conventional, and this is also corroborated by direct measurements of the adiabatic temperature-change \cite{Vives2011}. Significant cooling of the sample was measured when the stress was rapidly released. Values around 10 K for Cu-based alloys \cite{Rodriguez1980} and around 20 K for Ti-Ni based alloys have been reported \cite{Cui2012}.


In contrast to all other giant caloric materials, the transition entropy change is solely related to the change in structure at the transition (contributions from other degrees of freedom are negligeable). Indeed such an entropy-change is mostly vibrational and originates from low-lying transverse phonons in the TA$_2$ branch of the open cubic high-temperature phase (details can be found in \cite{Planes2001}).

\section{Magnetic Heusler alloys and the inverse magnetocaloric effect.}

Heusler alloys are $X_2YZ$ intermetallics with a $Fm3m$ structure where $X$ atoms occupy the 8c Wyckoff positions and $Y$ and $Z$ atoms the 4a and 4b positions, respectively. These alloys have received considerable interest for several decades \cite{Graf2011} which was further promoted with the report that Ni$_2$MnGa could exhibit large deformations by the application of a moderate magnetic field \cite{Ullakko1996}. At present, strains as large as 10 \% have been reported in off-stoichiometric compounds for fields below 1 T \cite{Sozinov2002}. The large strains are related to the martensitic transition taking place in these magnetic alloys. As a consequence of its lower symmetry, the martensitic phase exhibits a heterostructure formed by twin-related structural domains (variants) \cite{Battacharya2003} and magnetic domains. There is strong interplay between structural and magnetic degrees of freedom at the mesoscale of these domains. Owing to the high mobility of the twin boundaries and a high magnetic anisotropy, applying a magnetic field causes twin boundary motion that gives rise to these large field-induced deformations. Alloys exhibiting this property are known as magnetic shape memory alloys \cite{Soderberg2006}.

In the Ni-Mn-based Heusler family, Ni$_2$MnGa is the only ferromagnetic alloy that undergoes a martensitic transition at the stoichiometric composition, but almost any alloy of this family undergoes a martensitic transformation at appropriate off-stoichiometric compositions \cite{Entel2006}. Among them, the particularly interesting ones are those exhibiting magnetic superelastic behaviour \cite{Kainuma2006,Krenke2007}. They also exhibit large magnetic-field-induced strains. But in this case, the strains are not due to a rearrangement of structural domains but rather to the possibility of inducing the reverse martensitic transition on applying a magnetic field. In this case, large magnetocrystalline anisotropy is not required, but there must be a significant change in the magnetic moment between the high-temperature phase and the martensitic phase.

In general, Ni-Mn-based Heusler materials show a rich variety of magnetic behaviour \cite{Acet2011} with associated multifunctional properties, which include, in addition to the aforementioned magnetic shape memory, giant magneto-resistance, exchange bias, and giant caloric effects which will be discussed in the following.

\begin{figure}[h]
\centerline{\includegraphics[height=7cm]{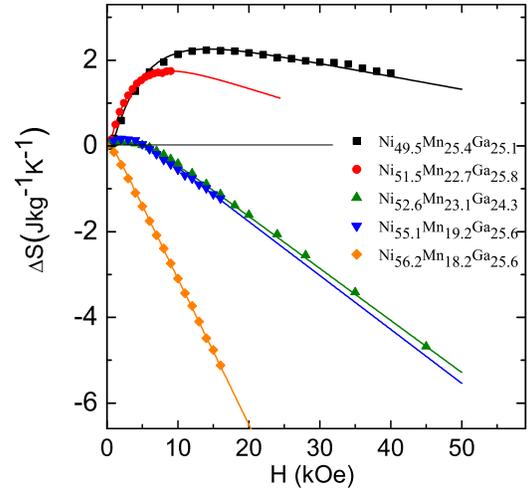}}
 \caption{Magnetocaloric effect in Ni-Mn-Ga. Average magnetic field induced entropy-change as a function of magnetic field for a family of composition related Ni-Mn-Ga alloys. Figure adapted from ref.\cite{Marcos2003}.}
 \label{figDSfieldNiMnGa}
\end{figure}

The magnetocaloric effect in Ni-Mn-Ga was first reported by Hu et al. \cite{Hu2000} who observed an increase in the entropy by applying a 1 T magnetic field. Soon after, the same authors reported in a sample with a slightly different composition an entropy decrease for fields above 1 T \cite{Hu2001}. Such puzzling behaviour was explained by Marcos et al. \cite{Marcos2002,Marcos2003b} who showed that it arises from the different length scales of the magnetostructural coupling. The inverse effect at low fields (entropy increase with magnetic field) is related to a magnetostructural coupling on the length scale of magnetic domains and martensitic variants: The strong uniaxial magnetocrystalline anisotropy of the martensitic phase results in a magnetic domain configuration with a lower magnetization than that of the high-temperature cubic phase, thus giving rise to an inverse magnetocaloric effect. At high fields, the coupling at a microscopic scale becomes dominant. Since for Ni-Mn-Ga the intrinsic magnetic moment in the martensite is larger than in the cubic phase, the magnetization increases at the transition and the magnetocaloric effect becomes conventional. Interestingly, when the composition is varied in such a way that the martensitic transition temperature approaches the Curie point, the magnetic anisotropy weakens with a corresponding decrease of the inverse contribution, and the conventional magnetocaloric effect becomes dominant. Actually, for Ni-Mn-Ga alloys, optimum magnetocaloric properties occur when both the martensitic and ferromagnetic transitions take place close to one another \cite{Pareti2003}. This behaviour is illustrated in fig. \ref{figDSfieldNiMnGa} which shows the magnetocaloric effect as a function of magnetic field for a series of Ni-Mn-Ga alloys in which the martensitic transition temperature approaches the Curie point as the Mn concentration decreases.

\begin{figure}[h]
\centerline{\includegraphics[height=35mm]{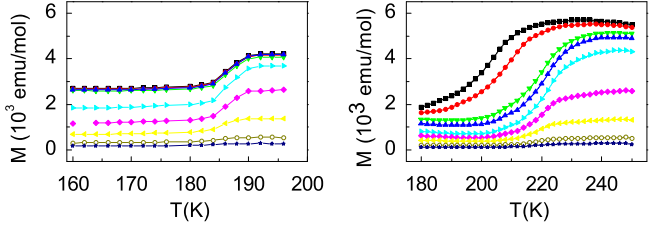}}
 \caption{Magnetization in Ni-Mn-Z magnetic shape memory alloys. Temperature dependence of the magnetization for  Ni$_{50}$Mn$_{35}$Sn$_{15}$ (left panel) and  Ni$_{50}$Mn$_{34}$In$_{16}$ (right panel) for selected magnetic fields. From bottom to top: H=0.1,0.2,0.5,1,2,10,50 kOe (left panel) and  H=0.1,0.5,2,5,10,20,30,40,50 kOe (right panel). Figure adapted from \cite{Planes2009}}
 \label{figMagnetizationNiMnZ}
\end{figure}

\begin{figure}[h]
\centerline{\includegraphics[height=7cm]{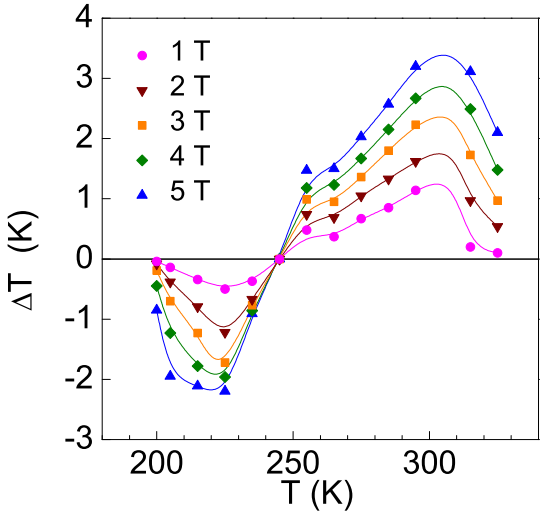}}
 \caption{Magnetocaloric effect in Ni-Mn-In. Temperature dependence of the adiabatic temperature-change for Ni$_{50}$Mn$_{34}$In$_{16}$. Figure adapted from \cite{Aksoy2007}.}
 \label{figDTDSNiMnIn}
\end{figure}

Ni-Mn-Z alloys with Z different from Ga behave significantly different from Ni-Mn-Ga. On the one hand, the magnetocrystalline anisotropy is low in both martensitic and cubic phases. On the other hand, the intrinsic magnetic moment of martensite is smaller than in the cubic phase. Such a decrease in the magnetic moment is a consequence of an enhancement of the antiferromagnetic correlations between Mn-atoms located at the 4b positions \cite{Aksoy2009}. Such a decrease is illustrated in fig. \ref{figMagnetizationNiMnZ} for Ni-Mn-Sn and Ni-Mn-In alloys. Consistent with the decrease in the magnetization, the martensitic transition shifts to lower temperatures with applied magnetic field, and consequently, the magnetocaloric effect is inverse in the vicinity of the martensitic transition, as firstly reported for Ni-Mn-Sn \cite{Krenke2005}. Later, other alloys from the Ni-Mn-Z family (with Z as In and Sb, and related quaternary alloys) were also reported to exhibit an inverse magnetocaloric effect \cite{Moya2007,Han2006,Khan2007,Krenke2007b}. At the Curie point, the alloys exhibit a conventional magnetocaloric effect where the entropy-change is solely due to a magnetic contribution. The inverse character of the magnetocaloric effect was confirmed by direct measurements of the adiabatic temperature-change \cite{Moya2007,Aksoy2007}. An example is shown in figure \ref{figDTDSNiMnIn}, which displays the measured adiabatic temperature change for Ni-Mn-In around the martensitic and magnetic transitions.

It is worth remarking that in Ni-Mn based Heusler alloys, the physical mechanism of the magnetocaloric effect is the same as that leading to their unique magnetic-field induced strain. On the one hand, the magnetostructural coupling at the mesoscale is responsible for the magnetic-shape memory and the low field inverse magnetocaloric effect in Ni-Mn-Ga alloys. Both effects require a large magnetocrystalline anisotropy in the martensitic state. On the other hand, the magnetostructural coupling at a microscopic scale gives rise to magnetic superelasticity and the inverse magnetocaloric effect. In this case a significant difference in the magnetic moment of the two phases is necessary in such a way that the martensitic transition can be driven by applying a magnetic field. The decresase in the magnetic moment is a consequence of a decrease in the distance between excess Mn-atoms in the martensitic unit cell compared to the cubic one which results in an enhancement of antiferromagnetic exchange \cite{Sokolovskiy2012}.

\begin{figure}[h]
\centerline{\includegraphics[height=7cm]{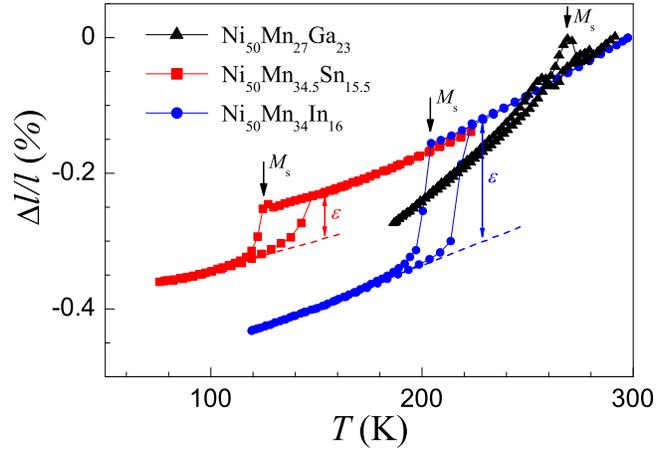}}
 \caption{Length change in magnetic shape memory alloys. Temperature dependence of the relative length change for Ni$_{50}$Mn$_{34}$In$_{16}$, Ni$_{50}$Mn$_{34.5}$Sn$_{15.5}$ and Ni$_{50}$Mn$_{27}$Ga$_{23}$ polycrystalline alloys. Figure adapted from \cite{Aksoy2007b}.}
 \label{figlengthchange}
\end{figure}

Typically, the volume change at the martensitic transition in shape memory alloys is negligibly small. Nevertheless, for some alloys of the Ni-Mn-Z family (Z different from Ga), a consequence of the strong interaction between magnetic and structural degrees of freedom is a difference between the unit cell volume of the martensitic phase and that of the cubic one. A first indication for such a volume change is evidenced by the relative length change at the temperature-induced martensitic transition (in the absence of any external magnetic field or stress) shown in figure \ref{figlengthchange} for several polycrystalline alloys \cite{Aksoy2007b}. To a good approximation, the relative volume-change amounts to three times the relative length-change. While for Ni-Mn-Ga the coupling at a microscopic level is small, and consequently the volume change at the transition is negligible, there is a noticeable volume change for the rest of alloys of the family with strong coupling. The martensitic transition in these alloys will be influenced by hydrostatic pressure \cite{Manosa2008} and they will be prone to exhibit a giant barocaloric effect.

\begin{figure}[h]
\centerline{\includegraphics[height=6cm]{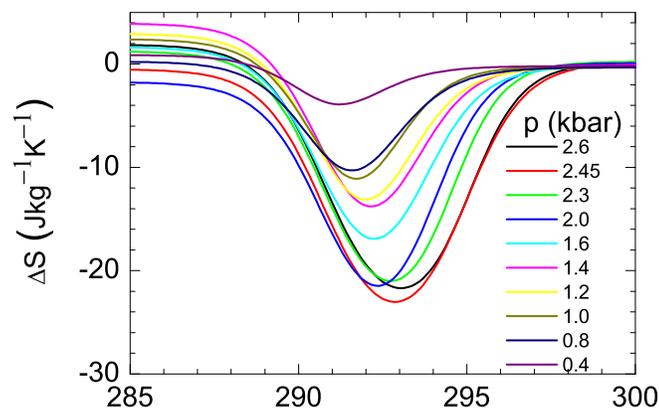}}
 \caption{Barocaloric effect in Ni-Mn-In. Temperature dependence of the pressure-induced entropy-change for Ni$_{49.26}$Mn$_{36.08}$In$_{14.66}$. Figure adapted from ref. \cite{Manosa2010}.}
 \label{figDSbarocaloricNiMnIn}
\end{figure}

The giant barocaloric effect was first reported for a Ni-Mn-In sample \cite{Manosa2010} from calorimetric measurements under hydrostatic pressure. Results are illustrated in figure \ref{figDSbarocaloricNiMnIn}, which shows the temperature-dependence of the pressure-induced entropy-change for selected values of hydrostatic pressures. It is found that the barocaloric effect in this compound is conventional. It is slightly larger than that found for Gd$_5$Si$_2$Ge$_2$ and significantly larger than the effect found for La-Fe-Co-Si. Furthermore, the values measured for moderated pressures are larger than the corresponding magnetocaloric values for fields around 1T.

As previously mentioned, the martensitic transition is mainly accomplished by a shear distortion, and consequently, it is very sensitive to applied uniaxial strain. A stress-induced first-order transition gives rise to an associated elastocaloric effect. Conventional elastocaloric effect has been found in magnetic shape-memory alloys \cite{SotoParra2010,CastilloVilla2011}. Nevertheless the strong brittleness of these alloys precludes the application of large stresses, and consequently, the associated elastocaloric effect remains at lower values.

Ni-Mn-based magnetic shape memory alloys exhibit conventional barocaloric and elastocaloric effects while the magnetocaloric effect is inverse (with the exception of Ni-Mn-Ga). This indicates that the main contribution to the transition entropy-change in these alloys is associated to the structural degrees of freedom. Actually, the origins are the same as in nonmagnetic martensitic materials: the cubic high-temperature phase has a larger vibrational entropy than the close-packed phase as a consequence of the low energy TA$_2$ phonon branch \cite{Moya2009}. On the other hand, the total transition entropy, which contains a structural and a magnetic contribution, decreases as the martensitic transition is pushed below the Curie point either by tuning composition \cite{Kustov2009} or by applying a magnetic field \cite{Emre2013}. This reflects that the magnetic contribution becomes more and more important and eventually can compensate the structural contribution resulting in a vanishing total entropy-change. This has been suggested to be at the origin of the interesting phenomenon of kinetic arrest \cite{Sharma2007,Ito2008} in magnetic shape memory alloys.

\section{BaTiO$_3$ and the electrocaloric effect.}

The electrocaloric effect refers to the temperature and entropy-changes occurring in polar materials when an electric field is applied. The effect has been known for many decades \cite{Kobeco1930} but it received little attention due to its small magnitude. Paralleling the development in other caloric effects, the research was triggered on realizing that giant effects were expected close to a paraelectric-ferroelectric phase transition. This was shown to occur in ceramic films \cite{Mischenko2006} and later in ferroelectric copolymers and ferroelectric relaxors \cite{Neese2008}.

Although thin films exhibit large electrocaloric effects since they can support large driving fields, their performances are significantly lower than those of bulk oxides for which the temperature and entropy-changes per unit of applied field are an order of magnitude higher \cite{Valant2012,Moya2012}.

The prototype ferroelectric material is BaTiO$_3$. At high-temperatures, it crystallizes in a cubic (perovskite) structure (space group $Pm3m$). On cooling, it undergoes a structural transition to a tetragonal structure (space group $P4/mmm$). On further cooling, the material transforms to an orthorhombic phase (space group $Amm2$) at around T= 270 K and to a rhombohedral phase (space group $R3m$) around T=200 K \cite{Kwei1993}. At the cubic to tetragonal transition, the Ti$^{4+}$ ions move relatively to the O$^{-2}$ octahedra resulting in spontaneous polarization along the $c$-axis. Such an electrosturctural paraelectric-ferroelectric phase transition is responsible for the significant electrocaloric effect in this oxide.

\begin{figure}[h]
\centerline{\includegraphics[height=7cm]{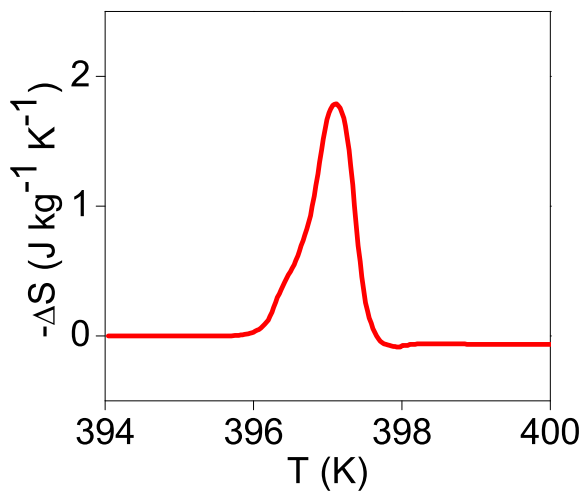}}
 \caption{Electrocaloric effect in BaTiO$_3$. Temperature dependence of the electric field induced entropy-change for BaTiO$_3$. Figure adapted from ref. \cite{Moya2012}.}
 \label{figDeltaSelectrocaloric}
\end{figure}

\begin{figure}[h]
\centerline{\includegraphics[height=8cm]{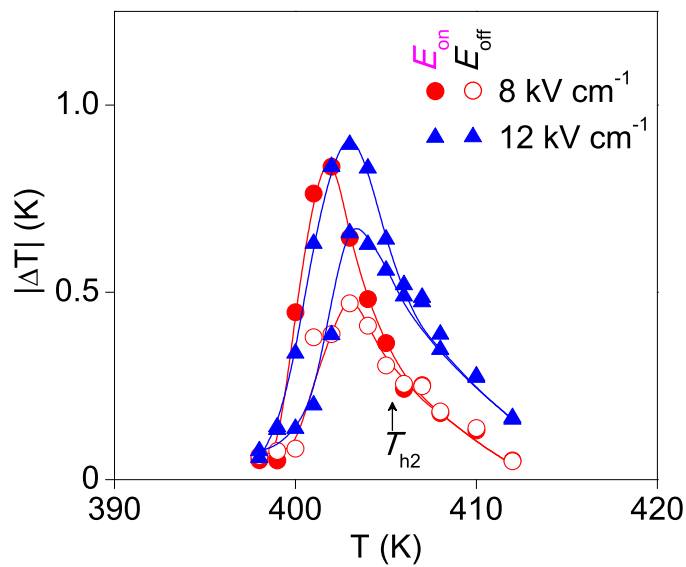}}
 \caption{Electrocaloric effect in BaTiO$_3$. Temperature dependence of the adiabatic temperature-change for electric fields of 8kVcm$^{-1}$ (circles) and 12 kVcm$^{-1}$ (triangles). Solid symbols correspond to the application of the field and open symbols, to the removal of the field. Figure adapted from ref. \cite{Moya2012}.}
 \label{figDeltaTelectrocaloric}
\end{figure}

The electrocaloric effect measured using calorimetry under electric field in a BaTiO$_3$ single crystal is illustrated in figure \ref{figDeltaSelectrocaloric} for an electric field of 3 kV cm$^{1}$. The peak value of the entropy-change coincides within experimental errors with the transition entropy-change which indicates that for this sample low values of the applied field are enough to achieve the total transition entropy-change. The electrocaloric effect is conventional, which is in agreement with the larger polarization of the tetragonal phase. Direct measurements of the adiabatic temperature-change, shown in figure \ref{figDeltaTelectrocaloric}, confirm this issue: the sample heats on applying an electric field. It is interesting to note that above a certain temperature (labeled T$_{h2}$ in the figure), the measured adiabatic temperature-change values are reversible. This behaviour is intimately related to the hysteresis of the first-order electrostructural transition in this sample \cite{Moya2012}.

The cubic to tetragonal phase transition encompasses a change in the volume of the unit cell of around 0.037 cm$^3$mol$^{-1}$ \cite{Samara1971}. The volume of the low-temperature tetragonal ferroelectric phase is larger than that of the cubic phase and therefore an inverse barocaloric effect is expected for this compound. Preliminary calorimetric measurements under hydrostatic pressure \cite{SternTaulats2013} do evidence such an inverse barocaloric effect. Interestingly, the maximum value for the entropy-change (around 2.5 J kg$^{-1}$K$^{-1}$) is already obtained for pressures as low as 1kbar.

\section{Conclusions and future prospects.}

\begin{center}
\begin{table*}[ht]
\small
\hfill{}
  \caption{\ Transition entropy change ($\Delta S_t$), measured  field-induced isothermal entropy ($\Delta S$) and adiabatic temperature ($\Delta T$) changes for several caloric effects obtained for the indicated values of the corresponding field. All data are in absolute value. (a) ref. \cite{Gschneidner2005}, (b) ref. \cite{Yuce2012}, (c) ref.\cite{Manosa2011}, (d) ref \cite{Bonnot2008}, (e) ref \cite{Moya2007}, (f) ref. \cite{Manosa2010} and (g) ref. \cite{Moya2012}.}
  \label{Taula1}
  \hfill{}
\begin{tabular}{llllll}
    \hline
        & Gd$_5$Si$_2$Ge$_2$ & LaFe$_{11.33}$Co$_{0.47}$Si$_{1.2}$ & Cu$_{68.13}$Zn$_{15.74}$Al$_{16.13}$ & Ni$_{49.26}$Mn$_{36.08}$In$_{14.66}$ & BaTiO$_3$  \\
    \hline
    $\Delta S_t$ (Jkg$^{-1}$K$^{-1}$) & 21.0 & 11.4 & 21.0 & 27.0 & 2.2 \\  
    \hline
    $\mu_0H$ (T) & 1 & 1 & - & 1 & - \\
    $\Delta S$ (Jkg$^{-1}$K$^{-1}$) & 8 & 2 & - & 10 & -  \\
    $\Delta T$ (K) & 4 & 1 & - & 0.5 & - \\
    \hline
    $p$ (kbar) & 1 & 1 & - & 1 & - \\
    $\Delta S$ (Jkg$^{-1}$K$^{-1}$) & 8 & 5 & - & 11 & -  \\
    $\Delta T$ (K) & $\sim$0.5 & 1.5 & - & - & - \\
    \hline
    $\sigma$ (MPa) & - & - & 5 & - & - \\
    $\Delta S$ (Jkg$^{-1}$K$^{-1}$) & - & - & 21 & - & -  \\
    $\Delta T$ (K) & - & - & 10 & - & - \\
    \hline
    $E$ (kVcm$^{-1}$) & - & - & - & - & $\sim$10 \\
    $\Delta S$ (Jkg$^{-1}$K$^{-1}$) & - & - & - & - & 2.2  \\
    $\Delta T$ (K) & - & - & - & - & 1 \\
    \hline
    Ref. & a,b & c,d & e & f & g \\
    \hline
  \end{tabular}
  \hfill{}
\end{table*}
\end{center}

In general, the different thermodynamic degrees of freedom can lead to several caloric effects. The key point for these effects to be large enough for potential practical applications is the occurrence of a phase transition. We have illustrated the effects reported so far in a variety of materials. The relevant quantities for fields readily accessible in practical applications are summarized in table \ref{Taula1}.

The magnitude of all these caloric effects open new perspectives for designing solid-state refrigeration devices as alternatives to the presently existing technologies.

In most materials, the hysteresis associated with the first-order phase transition represents a drawback because it reduces the refrigerant capacity. Most effort is devoted at finding materials where the hysteresis is small in comparison with the shift of the transition with field. The detrimental effect of hysteresis can also be reduced by taking advantage of the cross-response of several materials which enables using simultaneously more than one field.

The existence of the inverse caloric effects can also represent an added value, since devices with an appropriate combination of both conventional and inverse effects can enhance the refrigerant efficiency of a given device.

Recently efforts have been undertaken in order to deal with caloric properties from first principle calculation. These calculations are expected to give hints for the development of new materials with desired caloric properties or simply for optimizing caloric properties of already known materials. The validity of this point of view is supported by the fact that they  have proved to be able to nicely reproduce caloric behaviour of given materials \cite{deOliveira2013}. These investigations are approached with the idea of finding materials undergoing a phase transition involving a large change of the properties giving rise to the caloric effects of interest. There are, however, still few works with predictive character. By combining density functional modelling and Monte Carlo simulations Siewert et al \cite{Siewert2011} suggested that Pt-doping in Ni-Mn-based Heusler alloys improves the performances of these alloys, in particular their magnetocaloric properties \cite{Entel2013}. Furthermore, in  recent works, models based on first principles have been used to predict a giant electrocaloric effect in LiNbO$_3$ \cite{Rose2012} and both, electrocaloric \cite{Ponomareva2012} and elastocaloric \cite{Lisenkov2012} effects in Ba$_{0.55}$Sr$_{0.5}$TiO$_{3}$. Computational discovery of new caloric and multicaloric materials not only represents an important step towards the understanding of these properties but also opens a new route in relation to the development of this class of materials. While this certainly could yield to more efficient solid-state refrigeration devices, care must be taken since by using these numerical techniques it is not easy to have a reasonable estimation of hysteresis effects related to dissipative non-equilibrium aspects which reduce the refrigerant capacity.

Prototype refrigerators are, to a larger extent, based on the thermodynamic Brayton cycle, and the degree of field maturity is different for the several caloric effects: a significant number of prototypes using permanent magnets have already been developed in the case of the magnetocaloric effect (a good compilation can be found in \cite{Yu2010}), and very recently a refrigerator based on the electrocaloric effect has been reported \cite{Jia2012} which uses multilayer capacitors of BaTiO$_3$ \cite{Kar2010}. With regards to barocaloric and elastocaloric effects the research is still on the early stages.

While recent progress in the development and understanding of multicaloric materials have resulted in significant advances that suggest that solid-state refrigeration can become a reality in a near future, there are still important challenges to overcome both in the basic understanding and design of new materials and also on the engineering of prototypes before these technologies become commercially competitive.

\section{Acknowledgements}

We are grateful to our recent PhD students and post-docs J. Marcos, E. Bonnot, X. Moya, T. Krenke, E. Duman, S. Aksoy, D. Gonz\'alez-Comas, I. Titov, E. Stern-Taulats, D. Soto-Parra, S. Y\"uce and B. Emre for their collaboration on the topics covered in this paper. Financial support is acknowledged to CICyT (Spain) under project MAT2010-15114 and to Deutsche Forschungsgemeinschaft, Project SPP1239.

\providecommand*{\mcitethebibliography}{\thebibliography}
\csname @ifundefined\endcsname{endmcitethebibliography}
{\let\endmcitethebibliography\endthebibliography}{}


\end{document}